\documentclass[12pt,cite,epsf,epsfig]{article}
\usepackage{epsfig}
\usepackage[dvips]{color}

\begin{document}

\title{Maximal $CP$-Violation in Neutrino Mass Matrix in light of the latest Daya Bay result on $\theta_{13}$}
\author{Surender Verma\thanks{s\_7verma@yahoo.co.in}}
\date{\textit{Department of Physics, Government Degree College, Arki 173208, INDIA.}}
\maketitle
\begin{abstract}
The Daya Bay Collaboration\cite{1} has recently proclaimed discovery of non-zero reactor angle, $\theta_{13}$,  $\sin^{2}2\theta_{13}$=0.092$\pm$0.016(stat)$\pm$0.005(syst), at $5.2\sigma$ which is in conformity with the earlier observations of T2K\cite{2}, MINOS\cite{3} and DOUBLE CHOOZ\cite{4} last year. The discovery has immense implication for exploring $CP$-violation in the leptonic sector. In the present work, we examine $CP$-violation in Majorana neutrino mass matrix, in the basis where charged lepton mass matrix is diagonal, in which the ratio of elements are equal, also, termed as Strong Scaling Ans$\ddot{a}$tz (SSA). This Ans$\ddot{a}$tz has been known to explain the vanishing of $\theta_{13}$ and predict inverted hierarchy (IH) for the neutrino masses. However, to generate a non-zero value of $\theta_{13}$ consistent with Daya Bay result one has to deviate from SSA. This deviation will have important implications for $CP$-violation in the leptonic sector and is one of the issues addressed in the present work. $CP$ is maximally violated because, for central value of $\theta_{13}$ obtained from Daya Bay experiment, Dirac-type phase $|\delta|\approx \frac{\pi}{2}$, $\frac{3 \pi}{2}$ which corresponds to $|J_{CP}|\approx0.03$.  We have, also, studied imperative implications for Majorana-type phases $\alpha$, $\beta$ and effective neutrino mass, $M_{ee}$. The Majorana-type phases $\alpha$, $\beta$ are found to be sharply constrained.     
\end{abstract}

The latest observation of reactor angle, $\theta_{13}$, by the Daya Bay Collaboration\cite{1} has, finally, proved that there are three kinds of neutrino oscillations. The global fits of three mixing angles at $1\sigma$($3\sigma$) C.L. are\cite{5}

\begin{eqnarray}
\theta_{12}={34.0^{o}}^{+1.0^{o}(+2.9^{o})}_{-0.9^{o}(-2.7^{o})},
\theta_{23}={46.1^{o}}^{+3.5^{o}(+7.0^{o})}_{-4.0^{o}(-7.5^{o})},
\theta_{13}={7.27^{o}}^{+1.65^{o}(+4.12^{o})}_{-1.53^{o}(-5.45^{o})}.
\end{eqnarray}
The Daya Bay Collaboration\cite{1} provide, at $5.2\sigma$ C.L., the value of mixing angle $\theta_{13}$ given by
\begin{equation}
\theta_{13}={8.828^{o}}\pm 0.793 (stat)\pm 0.248 (syst).
\end{equation}
Recently, there have been numerous theoretical and phenomenological attempts to explain the observed non-zero value of $\theta_{13}$ and to explore $CP$-violation in the leptonic sector\cite{6}. Motivated by the recent measurement of $\theta_{13}$, it will be of immense importance to investigate $CP$-violation in the leptonic sector seeing that the evidence of non-zero $\theta_{13}$ yields to a potentially measurable Dirac-type $CP$ phase, $\delta$, in future neutrino oscillation experiments. The present work is focussed on the issue of $CP$-violation in Majorana neutrino mass matrix, $M_{\nu}$, with ratio of its elements being equal\cite{7} and is given by

\begin{equation}
M_{\nu}=\left( \begin{array}{ccc}
M_{ee} & M_{e\mu} & M_{e\tau} \\ 
M_{e\mu} & M_{\mu\mu} & M_{\mu \tau}\\ 
M_{e\tau} & M_{\mu \tau} & M_{\tau \tau}
\end{array}\right)\equiv\left( \begin{array}{ccc}
A & B & B/s \\ 
B & D & D/s \\ 
B/s & D/s & D/s^{2}
\end{array}\right). 
\end{equation}  
The Majorana neutrino mass matrix, $M_{\nu}$ (Eqn. (3)) has rank 2 and predict inverted hierarchy (IH) for the neutrino masses with $m_{3}=0$ and $\theta_{13}=0$ indicating that there will not be any $CP$-violation in neutrino oscillation experiments. Therefore, in vista of Daya Bay results, one has to deviate from SSA to accommodate the non-zero value of $\theta_{13}=8.828^{o}$ measured in Daya Bay experiment and to get a correct phenomenology. The Seesaw\cite{8} realization of SSA and its implications for thermal leptogenesis have been discussed in Ref.\cite{9}. The form (Eqn. (3)) necessarily occurs regardless of the form of right handed Majorana neutrino mass matrix, $M_{R}$, if third row of Dirac neutrino mass matrix, $M_{D}$, multiplied with $s$ is equal to the second row\cite{7}. The structure of Majorana neutrino mass matrix discussed in the present work has the additional virtue that it is stable against RG effects\cite{7, 10}.   Out of three possible independent scale invariant ways\cite{7} only one (Eqn. (3)) is consistent with the current data on neutrino masses and mixings\cite{7}. We establish deviations from SSA by introducing a scale breaking parameter $\kappa$ in Majorana neutrino mass matrix and is given by

\begin{equation}
M_{\nu}^{\prime}=\left( \begin{array}{ccc}
M_{ee} & M_{e\mu} & M_{e\tau} \\ 
M_{e\mu} & M_{\mu\mu} & M_{\mu \tau}\\ 
M_{e\tau} & M_{\mu \tau} & M_{\tau \tau}
\end{array}\right)^{\prime}\equiv\left( \begin{array}{ccc}
A & B & B/s \\ 
B & D & D/s \\ 
B/s & D/s & D/(1+\kappa)s^{2}
\end{array}\right)
\end{equation} 
where the parameters $A, B$ and $D$ are complex and $s$, $\kappa$ are assumed to be real. To discuss the phenomenology of the model it is useful to note that Eqn. (4) yields three constraining equations which will be used to locate the parameter space of the model allowed by existing data on neutrino masses and mixing angles.

\begin{equation}
m_1 x+\tilde{m_2} y+\tilde{m_3} z=0,
\end{equation}
\begin{equation}
m_1 x^{\prime}+\tilde{m_2} y^{\prime}+\tilde{m_3} z^{\prime}=0,
\end{equation}
\begin{equation}
m_1 x^{\prime \prime}+\tilde{m_2} y^{\prime \prime}+\tilde{m_3} z^{\prime \prime}=0,
\end{equation}
where $\tilde{m_2}\equiv m_2 e^{2 i \alpha}$, $\tilde{m_3}\equiv m_3 e^{2 i (\beta+\delta)}$ and $m_1$, $m_2$ and $m_3$ are neutrino masses. The complex parameters $x, y, z, x^{\prime}, y^{\prime},z^{\prime}, x^{\prime \prime}, y^{\prime \prime}$ and $z^{\prime \prime}$ are given by
   
\begin{equation}
 \left.\begin{array}{c}
x = U_{e1} (U_{\mu 1} - s U_{\tau 1})\\
y = U_{e2} (U_{\mu 2} - s U_{\tau 2})\\
z = U_{e3} (U_{\mu 3} - s U_{\tau 3})
\end{array}  \right\},
\end{equation}

\begin{equation}
 \left.\begin{array}{c}
x^{\prime} = U_{\mu 1} (U_{\mu 1} - s U_{\tau 1})\\
y^{\prime} = U_{\mu 2} (U_{\mu 2} - s U_{\tau 2})\\
z^{\prime} = U_{\mu 3} (U_{\mu 3} - s U_{\tau 3})
\end{array}  \right\},
\end{equation}

\begin{equation}
 \left.\begin{array}{c}
x^{\prime \prime} = U_{\tau 1} (U_{\mu 1} - s (1+\kappa) U_{\tau 1})\\
y^{\prime \prime} = U_{\tau 2} (U_{\mu 2} - s (1+\kappa) U_{\tau 2})\\
z^{\prime \prime} = U_{\tau 3} (U_{\mu 3} - s (1+\kappa) U_{\tau 3})
\end{array}  \right\},
\end{equation}
with 

\begin{equation}
U= \left(
\begin{array}{ccc}
U_{e1} & U_{e2} & U_{e3} \\
U_{\mu 1} & U_{\mu 2} & U_{\mu 3} \\
U_{\tau 1} & U_{\tau 2}& U_{\tau 3}
\end{array}
\right), \nonumber
\end{equation}
\begin{equation}
\equiv \left(
\begin{array}{ccc}
c_{12}c_{13} & s_{12}c_{13} & s_{13}e^{-i\delta} \\
-s_{12}c_{23}-c_{12}s_{23}s_{13}e^{i\delta} &
c_{12}c_{23}-s_{12}s_{23}s_{13}e^{i\delta} & s_{23}c_{13} \\
s_{12}s_{23}-c_{12}c_{23}s_{13}e^{i\delta} &
-c_{12}s_{23}-s_{12}c_{23}s_{13}e^{i\delta} & c_{23}c_{13}
\end{array}
\right)\left(
\begin{array}{ccc}
1 & 0 & 0 \\ 0 & e^{i\alpha} & 0 \\ 0 & 0 & e^{i(\beta+\delta)}
\end{array}
\right),
\end{equation}
 $c_{ij}\equiv\cos{\theta_{ij}}$ and $s_{ij}\equiv\sin{\theta_{ij}}$\cite{11}. Eqns. (5-7) can be written in terms of the ratios $\frac{\tilde{m_2}}{m_1}$, $\frac{\tilde{m_3}}{m_1}$ as
  
\begin{equation}
x+\frac{\tilde{m_2}}{m_1} y+\frac{\tilde{m_3}}{m_1} z=0,
\end{equation}
\begin{equation}
 x^{\prime}+\frac{\tilde{m_2}}{m_1} y^{\prime}+\frac{\tilde{m_3}}{m_1} z^{\prime}=0,
\end{equation}
\begin{equation}
 x^{\prime \prime}+\frac{\tilde{m_2}}{m_1} y^{\prime \prime}+\frac{\tilde{m_3}}{m_1} z^{\prime \prime}=0.
\end{equation}

Using Eqns. (13-14) the mass ratios $\frac{\tilde{m_2}}{m_1}$ and $\frac{\tilde{m_3}}{m_1}$ can be written as

\begin{eqnarray}
\frac{\tilde{m_2}}{m_1}=-\frac{\left(c_{23} s_{12} s_{13}+c_{12} s_{23} e^{i \delta }\right) \left(s_{12} \left(c_{23}+s s_{23}\right)-c_{12} s_{13} e^{i
   \delta } \left(c_{23} s-s_{23}\right)\right)}{\left(c_{12} c_{23} s_{13}-s_{12} s_{23} e^{i \delta }\right) \left(c_{12}
   \left(c_{23}+s s_{23}\right)+s_{12} s_{13} e^{i \delta } \left(c_{23} s-s_{23}\right)\right)},
\end{eqnarray}
\begin{eqnarray}
\frac{\tilde{m_3}}{m_1}=\frac{c_{23} e^{i \delta } \left(s_{12} \left(c_{23}+s s_{23}\right)-c_{12} s_{13} e^{i \delta } \left(c_{23}
   s-s_{23}\right)\right)}{\left(c_{23} s-s_{23}\right) \left(-c_{12} c_{23} s_{13}+s_{12} s_{23} e^{i \delta }\right)}.
\end{eqnarray}

The Taylor series expansion of mass ratios $\frac{m_2}{m_1}$, $\frac{m_3}{m_1}$ and the Majorana phases $\alpha$, $\beta$ upto first order in $s_{13}$ can be written as

\begin{eqnarray}
\frac{m_2}{m_1}=\left|\frac{\tilde{m_2}}{m_1}\right|\approx \left|1+\frac{s_{13} e^{-i \delta } \left(c_{23} \left(c_{23}+s s_{23}\right)+s_{23} e^{2 i \delta } \left(s_{23}-c_{23}
   s\right)\right)}{c_{12} s_{12} s_{23} \left(c_{23}+s s_{23}\right)}\right|,
\end{eqnarray}
\begin{eqnarray}
\frac{m_3}{m_1}=\left|\frac{\tilde{m_3}}{m_1}\right|\approx\left|\frac{c_{23} \left(c_{23}+s s_{23}\right)}{s_{23} \left(c_{23} s-s_{23}\right)}+\frac{c_{12} c_{23} s_{13} e^{-i \delta }
   \left(c_{23} \left(c_{23}+s s_{23}\right)+s_{23} e^{2 i \delta } \left(s_{23}-c_{23} s\right)\right)}{s_{12} s_{23}^2
   \left(c_{23} s-s_{23}\right)}\right|,
\end{eqnarray}

\begin{eqnarray}
\alpha\approx \frac{1}{2}Arg\left(1+\frac{s_{13} e^{-i \delta } \left(c_{23} \left(c_{23}+s s_{23}\right)+s_{23} e^{2 i \delta } \left(s_{23}-c_{23}
   s\right)\right)}{c_{12} s_{12} s_{23} \left(c_{23}+s s_{23}\right)}\right),
\end{eqnarray}
\begin{eqnarray}
\beta\approx\frac{1}{2}\left(Arg\left(\frac{c_{23} \left(c_{23}+s s_{23}\right)}{s_{23} \left(c_{23} s-s_{23}\right)}+\frac{c_{12} c_{23} s_{13} e^{-i \delta }
   \left(c_{23} \left(c_{23}+s s_{23}\right)+s_{23} e^{2 i \delta } \left(s_{23}-c_{23} s\right)\right)}{s_{12} s_{23}^2
   \left(c_{23} s-s_{23}\right)}\right)+\delta\right).
\end{eqnarray}

Substituting the values of mass ratios $\frac{\tilde{m_2}}{m_1}$ and $\frac{\tilde{m_3}}{m_1}$ from Eqns. (18-19) in Eqn. (15) we obtain

\begin{eqnarray}
-c_{23}s_{12}^{2}s_{23}-(1+\kappa)s s_{12}^{2}s_{23}^{2}-\frac{c_{23}^{2}(c_{23}(1+\kappa)s-s_{23})(c_{23}+s s_{23})}{(c_{23} s-s_{23})s_{23}}
-c_{12}^2s_{23}(c_{23}+(1+\kappa)s s_{23})\nonumber \\-\frac{c_{12}s_{13}e^{-i \delta}(c_{23}^{2}(1+\kappa)s+c_{23}(-1+(1+\kappa) s^{2})s_{23})}{s_{12}(c_{23} s-s_{23})s_{23}^{2}(c_{23}+s s_{23})}\nonumber \\
+c_{12}s_{13}e^{-i \delta}(\frac{(1+\kappa)s s_{23}^{2})(s_{23}e^{2 i \delta}(-c_{23} s+s_{23})+c_{23}(c_{23}+s s_{23}))}{s_{12}(c_{23} s-s_{23})s_{23}^{2}(c_{23}+s s_{23})}=0.
\end{eqnarray}
The real and complex part of Eqn. (22) can be written as
\begin{eqnarray}
c_{23}s_{12}s_{23}(c_{23}^{2}-\kappa s-c_{23}s_{23}-(1+\kappa)s s_{23}^{2})+c_{12}s_{13}(c_{23}^{2}\kappa s-c_{23}s_{23}-\kappa s s_{23}^{2})\cos{\delta}=0,
\end{eqnarray} 
\begin{eqnarray}
c_{12}s_{13}(c_{23}^{4}\kappa s-c_{23}^{3}s_{23}-2 c_{23}^{2}(1+\kappa)s s_{23}^{2}+c_{23}s_{23}^{3}+\kappa s s_{23}^{4})\sin{\delta}=0.
\end{eqnarray}
Solving Eqn. (23) and (24), for $s_{13}$ and $\kappa$ we obtain

\begin{eqnarray}
s_{13}=-\frac{s_{12}s_{23}}{2 c_{12}c_{23}\cos{\delta}}
\end{eqnarray}  
and
\begin{eqnarray}
\kappa=\frac{c_{23}s_{23}(c_{23}^{2}-s_{23}^{2}+2 s c_{23}s_{23})}{s (c_{23}^{2}-s_{23}^{2})^{2}}
\end{eqnarray}
Let us discuss the phenomenology of the Eqns. (18) and (25) to comprehend the interesting features of the correlation plots. For Eqn. (18) to be compatible with the solar neutrino mass hierarchy, the mass ratio $\frac{m_2}{m_1}$ should be greater than 1. If we put $\delta=\pi$ in Eqn. (18) we obtain
\begin{eqnarray}
\frac{m_2}{m_1}\approx1-\frac{s_{13}}{c_{12}s_{12}s_{23}(c_{23}+s s_{23})}
\end{eqnarray}
which does not satisfy solar neutrino mass hierarchy. Also, $m_2=m_1$ if $s_{13}=0$ in Eqn. (18). Thus, the points $\delta=\pi$ and $s_{13}=0$ are disallowed in neutrino mass models of the type given by Eqn. (4). In Fig. (1) we have plotted $R\equiv\frac{m_2}{m_1}$ as a function of $(s_{13}$, $\delta)$ and depict the parameter space (shaded region) for which solar neutrino mass hierarchy is satisfied. Furthermore, it is implicit from Eqn. (25) that the Dirac-type $CP$ violating phase $\delta$ should be in the II and III quadrant so as to make $s_{13}$ positive definite i.e. $90^o<\delta<270^o$. Thus, in light of the latest Daya Bay result on $\theta_{13}$, Majorana neutrino mass matrix (Eqn. (4)) is necessarily $CP$ violating (because $s_{13}\neq0$ and $\delta\neq\pi$ implying $J_{CP}\neq0$).

It is to be noted that we have not used these mass ratios (Eqns. (18-19)) and Majorana phases (Eqns. (20-21)) in our analysis which is
completely based on the exact Eqns. (13-15). They are given here for the
sake of illustration and to comprehend the interesting features of the correlation plots.

In our numerical analysis, we solve Eqns. (13) and (14) for two mass ratios $\frac{m_2}{m_1}$, $\frac{m_3}{m_1}$ and two Majorana phases $\alpha$, $\beta$ as a function of the other parameters of the model. All the known parameters are normally distributed with central values and errors given in Table 1. However, the parameters $s$ and $\kappa$, which are the free parameters of the model, are uniformly distributed. Substituting the values of $\frac{m_2}{m_1}$, $\frac{m_3}{m_1}$ and two Majorana phases $\alpha$, $\beta$ in Eqn. (15), we look for the allowed parameter space for which Eqn. (15) is satisfied.  We have illustrated the predictions of the model as correlations plots in Figs. (2-3) amongst various parameters of the model. To obtain the prediction of the model for $\theta_{13}$ we initially assume $\theta_{13}$ as free parameter giving full variation upto CHOOZ upper bound of $12^{o}$ (left panel of Fig. (2) and Fig. (3)). The right panel of Fig. (2) depicts the impact of Daya Bay observation of non-zero $\theta_{13}$, on the allowed parameter space and $CP$-violation. The strength of $CP$-violation in neutrino oscillations is described by the Jarlskog rephasing invariant\cite{12} given by

\begin{equation}
J_{CP}=\Im{\left[U_{e2}U_{\mu 3}U^{*}_{e3}U^{*}_{\mu2}\right]}
=c_{12}s_{12}c_{13}^2s_{13}c_{23}s_{23}\sin{\delta}
\end{equation}

In Fig. (2) we have depicted the correlation plots of $\kappa$, $\theta_{13}$ and $J_{CP}$. Fig (2(a)) has, previously, been obtained in Ref.\cite{7}. However, here we want to point out that to generate a non-zero value of $\theta_{13}$ consistent with Daya Bay result requires a sizeable breaking of $SSA$ which in Ref.\cite{7} is termed as ``weak scaling". Thus, present data on neutrino masses and mixings allows only weak scaling in Majorana neutrino  mass matrix. The correlation plots in the left column of figures have been obtained  assuming $\theta_{13}$ as a free parameter, however, correlation plots in the right column of the figures shows the predictions of the model with $\theta_{13}$ given by Daya Bay experiment. In the second row of Fig. (2) the prediction for $J_{CP}$ is shown as a function of the reactor angle, $\theta_{13}$.  It is implicit from the Fig. (2(d)) that the Majorana neutrino mass matrix model (Eqn. (4)) is necessarily $CP$-violating. Furthermore, it is rather interesting to find that $CP$ is maximally violated as for best fit value of $\theta_{13}$ (Eqn. (2)), $|J_{CP}|\approx0.03$ which corresponds to $|\delta|\approx \frac{\pi}{2}$, $\frac{3 \pi}{2}$ which is large enough to be measured in future oscillation experiments.

Neutrinoless double beta decay ($0\nu\beta\beta$)\cite{13} is an important low energy process which if observed, will establish the Majorana nature of neutrinos. The rate of this process is proportional to the modulus of (1,1) element of the effective Majorana neutrino mass matrix and is given by

\begin{equation}
M_{ee}=|\sum_{i}m_{i} U_{ei}^{2}|
\end{equation}
which in terms of the mass ratios $\frac{\tilde{m_2}}{m_1}$ and $\frac{\tilde{m_3}}{m_1}$ can be written as
\begin{eqnarray}
M_{ee}\approx \sqrt{\frac{\Delta m^2_{21}}{\left|\frac{\tilde{m_{2}}}{m_{1}}\right|^2-1}}\left(U_{e1}^{2}+\frac{\tilde{m_{2}}}{m_{1}}U_{e2}^{2}+\frac{\tilde{m_{3}}}{m_{1}}U_{e3}^{2}\right).
\end{eqnarray}
 In Fig. (3), we have shown the correlation plots of Majorana phases $\theta_{13}$,  $\alpha$, $\beta$ and effective Majorana neutrino mass $M_{ee}$. It is clear from Fig. (3(b)) that Majorana phase $\alpha$ ($\beta$) is sharply constrained to very narrow regions around $-50^o$ and $50^o$ ($-167^o$ and $-197^o$). We have, also, depicted correlation plot of $M_{ee}$ with $\theta_{13}$ in Fig. 3 (c) (with $\theta_{13}$ as free parameter) and 2(d) (with $\theta_{13}={8.828^{o}}\pm 0.793 (stat)\pm 0.248 (syst)$). We obtain $M_{ee}=$($0.046-0.066$)eV which is within the range of future $0\nu\beta\beta$ experiments probing $M_{ee}$ at the level of $10$meV to $50$meV\cite{14, 15, 16, 17}. A non-observance of $0\nu\beta\beta$ decay down to the level of $0.046$eV, which is achievable in future $0\nu\beta\beta$ experiments will rule out the Majorana neutrino mass model given by Eqn. (4). 
 
 In Conclusion, the experimental observations of non-zero $\theta_{13}$ by T2K\cite{2} and, recently, by Daya Bay\cite{1} experiment have triggered assiduous efforts to explore $CP$-violation in the leptonic sector. The Majorana neutrino mass matrix contain nine physical parameters and not all of them are measured by the neutrino experiments.  Moreover, in Majorana neutrino mass matrix (Eqn. (4)), there are two more parameters $s$ and $\kappa$ making it arduous to solve analytically. In view of this, numerical approach to investigate the phenomenology of neutrino mass matrix is very imperative. In the present work, we have explored the $CP$-violation in the neutrino sector where the ratios of elements of Majorana neutrino mass matrix is equal (Eqn. (4)) and is, also, known as Strong Scaling Ans$\ddot{a}$tz (SSA). We find that to obtain a value of $\theta_{13}$, consistent with recent result of Daya Bay experiment for $\theta_{13}$, require sizeable breaking of SSA. Thus, only weak scaling is allowed by current data on neutrino masses and mixings. Also, it is implicit from Fig. 2(d) that Majorana neutrino mass matrix (Eqn. (4)) is necessarily $CP$-violating. Furthermore, it is very interesting to find that $CP$ is maximally violated as for best fit value of $\theta_{13}$ (Eqn. (2)), $|J_{CP}|\approx0.03$ which corresponds to $|\delta|\approx \frac{\pi}{2}$, $\frac{3 \pi}{2}$ and is large enough to be measured in future oscillation experiments. Majorana phase $\alpha$ ($\beta$) is sharply constrained to very narrow regions around $-50^o$ and $50^o$ ($-167^o$ and $-197^o$). We obtain $M_{ee}=$($0.046-0.066$)eV which is within the range of future $0\nu\beta\beta$ experiments probing $M_{ee}$ at the level of $10$meV to $50$meV\cite{14, 15, 16, 17}. A non-observance of $0\nu\beta\beta$ decay down to the level of $0.046$eV, which is achievable in future $0\nu\beta\beta$ experiments will rule out the Majorana neutrino mass model given by Eqn. (4). 
      
\textit{\large{\textbf{Acknowledgement}}}\\
We thank University Grants Commission (UGC) and Ministry of Human Resource Development (MHRD) , Government of India, for providing de rigueur research facility in the college through various financial grants and programs like N-LIST and NMEICT.

\newpage
 \begin{table} [t]
\begin{center}
\begin{tabular}{cc}
  \hline\hline
    Parameters & Best fit $\pm 1\sigma$ ($\pm 3\sigma$) \\ \hline\hline
  $\Delta m^2_{21} [10^{-5} eV^{2}]$ & $7.59_{{-0.18} (-0.50)}^{{+0.20}(+0.60)}$ \\
  $|\Delta m^2_{31}| [10^{-3} eV^{2}]$ & $2.40_{{-0.09} (-0.27)}^{{+0.08} (+0.27)}$\\
  $\theta_{12}$ & ${34.0^{o}}^{+1.0^{o}(+2.9^{o})}_{-0.9^{o}(-2.7^{o})}$  \\
  $\theta_{23}$ & ${46.1^{o}}^{+3.5^{o}(+7.0^{o})}_{-4.0^{o}(-7.5^{o})}$ \\
  $\theta_{13}$ & ${7.27^{o}}^{+1.65^{o}(+4.12^{o})}_{-1.53^{o}(-5.45^{o})}$ \\
  \hline
\end{tabular}
\caption{The global fit of the neutrino mixing angles and mass-squared differences at $1 \sigma$ ($3 \sigma$) confidence levels.}
\end{center}
\end{table}

 \begin{figure}
\begin{center}
\epsfig{file=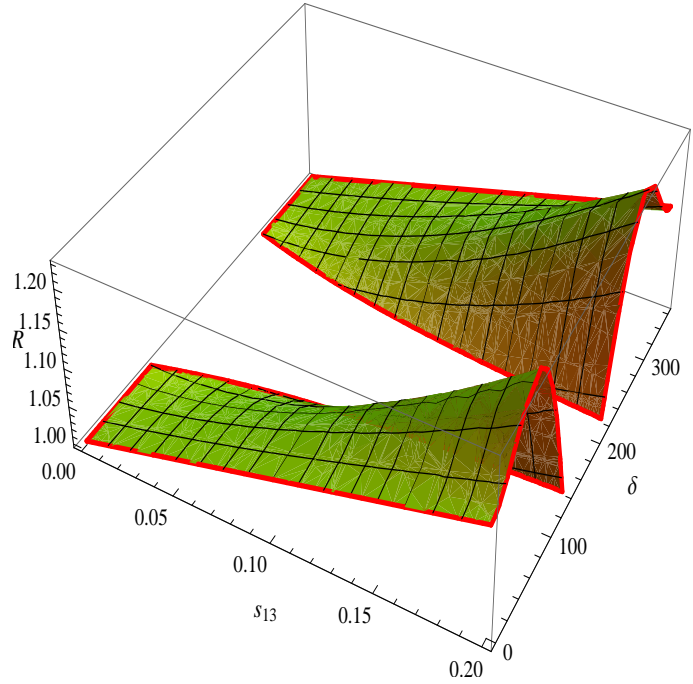, width=9.0cm, height=9.0cm} 
\caption{$R_{1}\equiv\frac{m_2}{m_1}$ as a function of $s_{13}$ and Dirac-Type $CP$ violating phase $\delta$.}
 \end{center}
\end{figure}
\begin{figure}
\begin{center}
{\epsfig{file=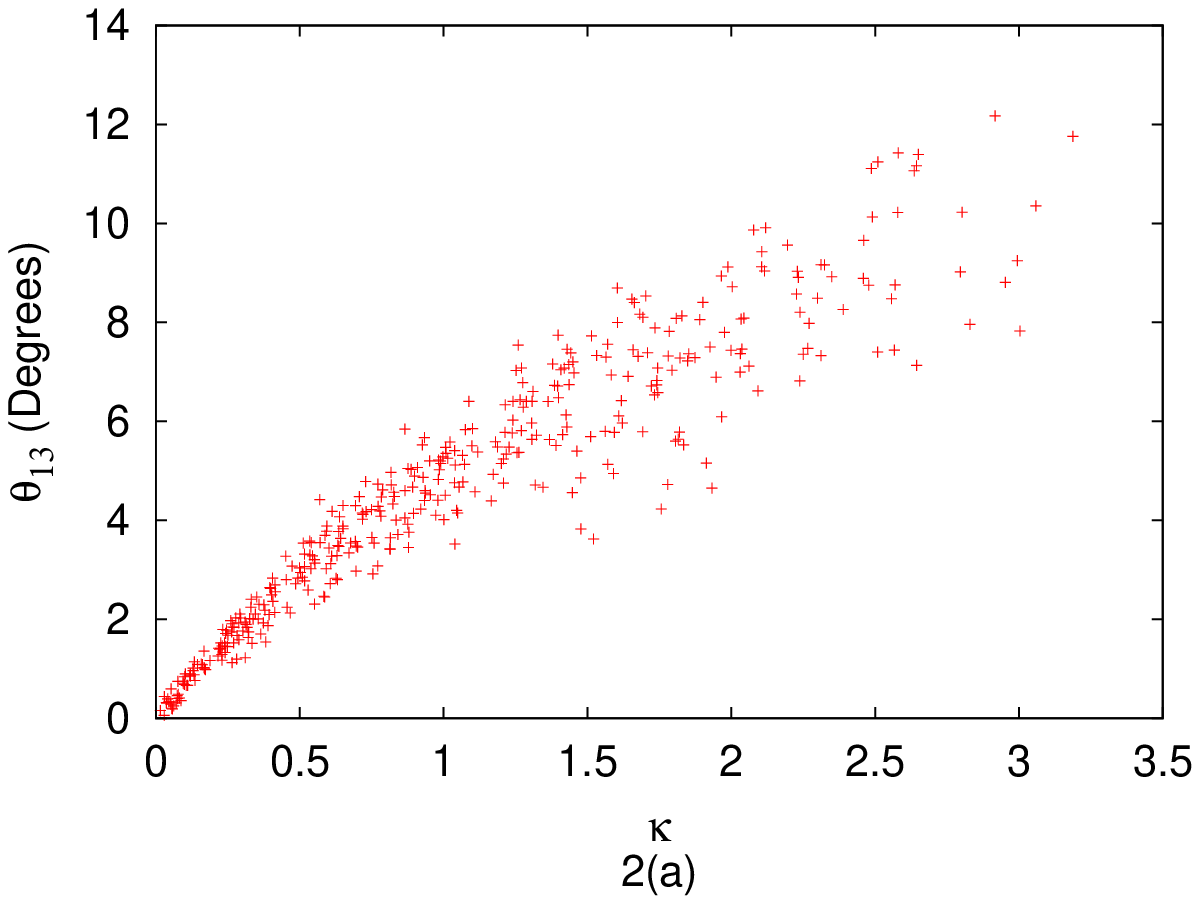, width=7.0cm, height=7.0cm} 
 \epsfig{file=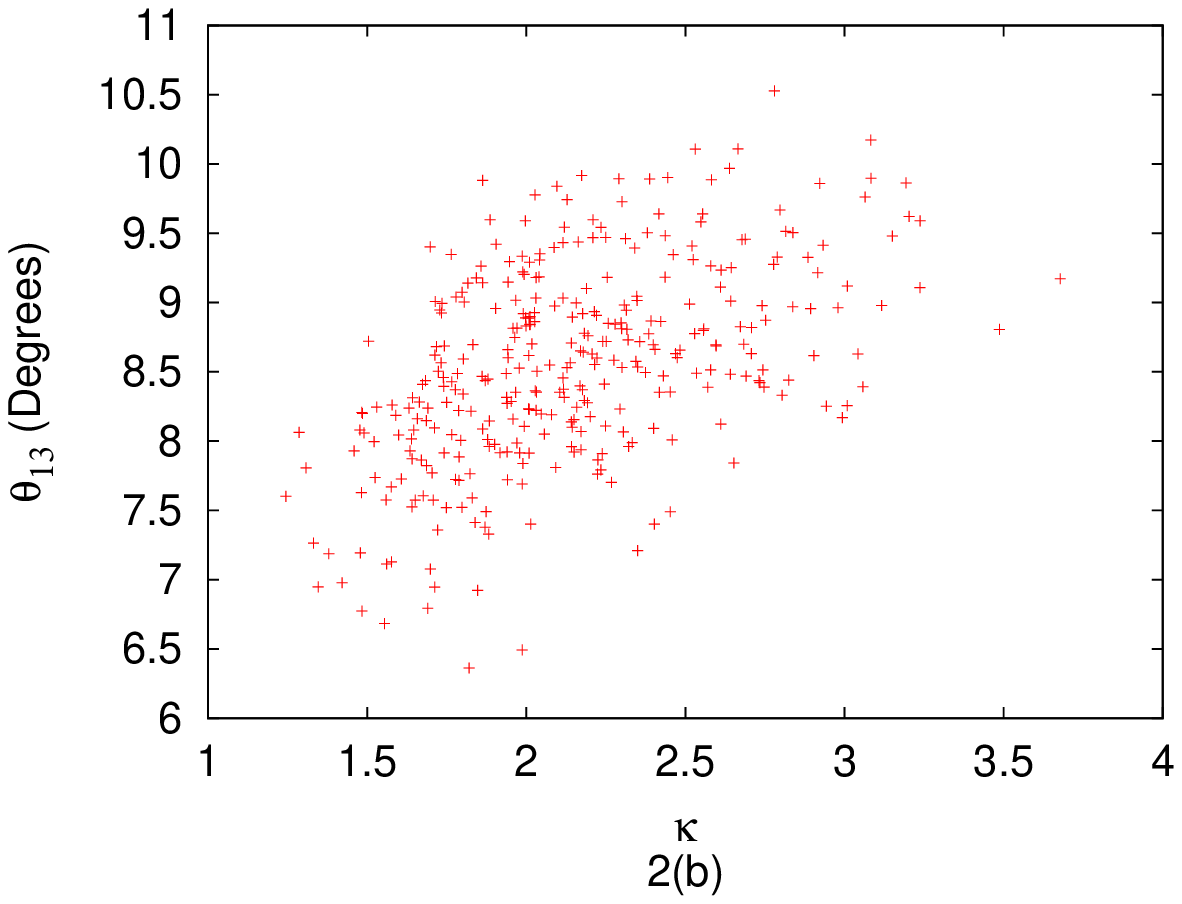, width=7.0cm, height=7.0cm}}
  {\epsfig{file=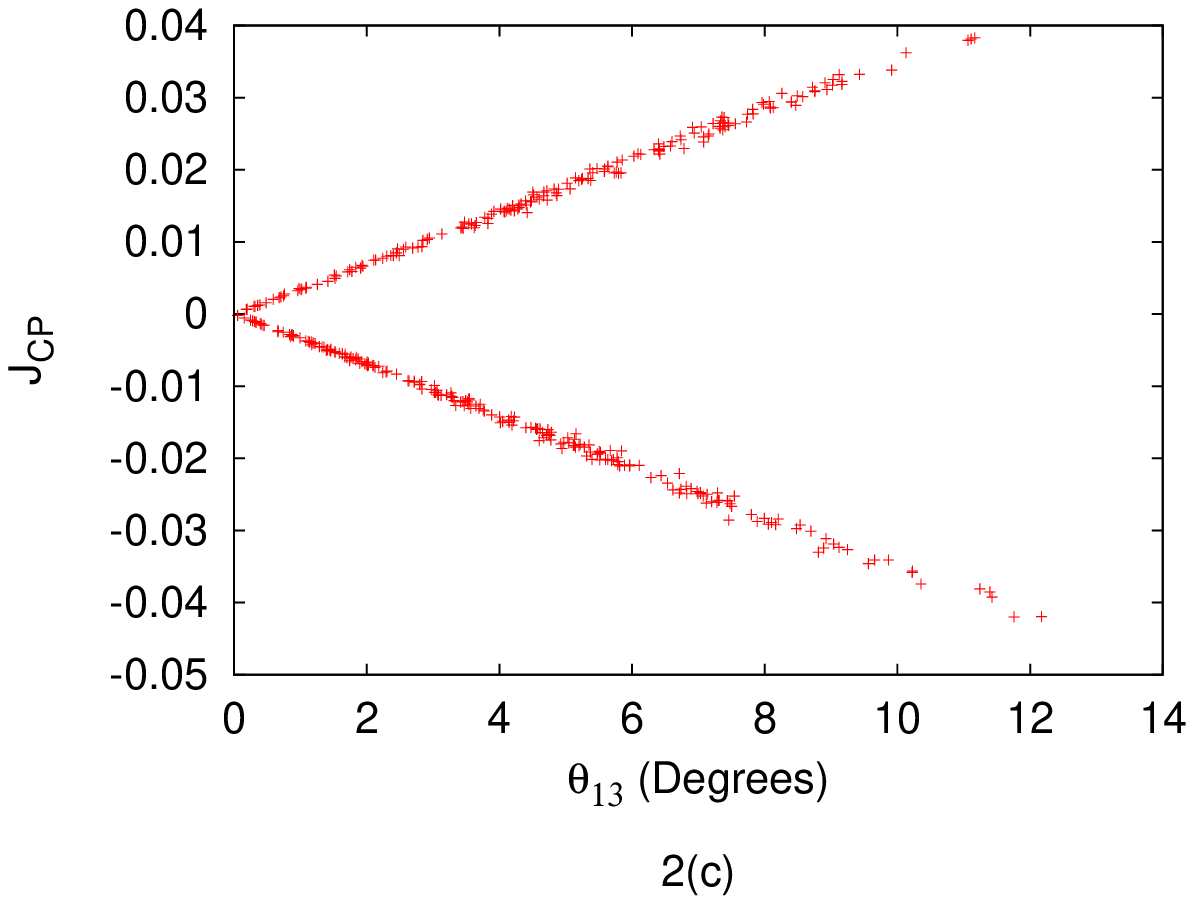, width=7.0cm, height=7.0cm} 
 \epsfig{file=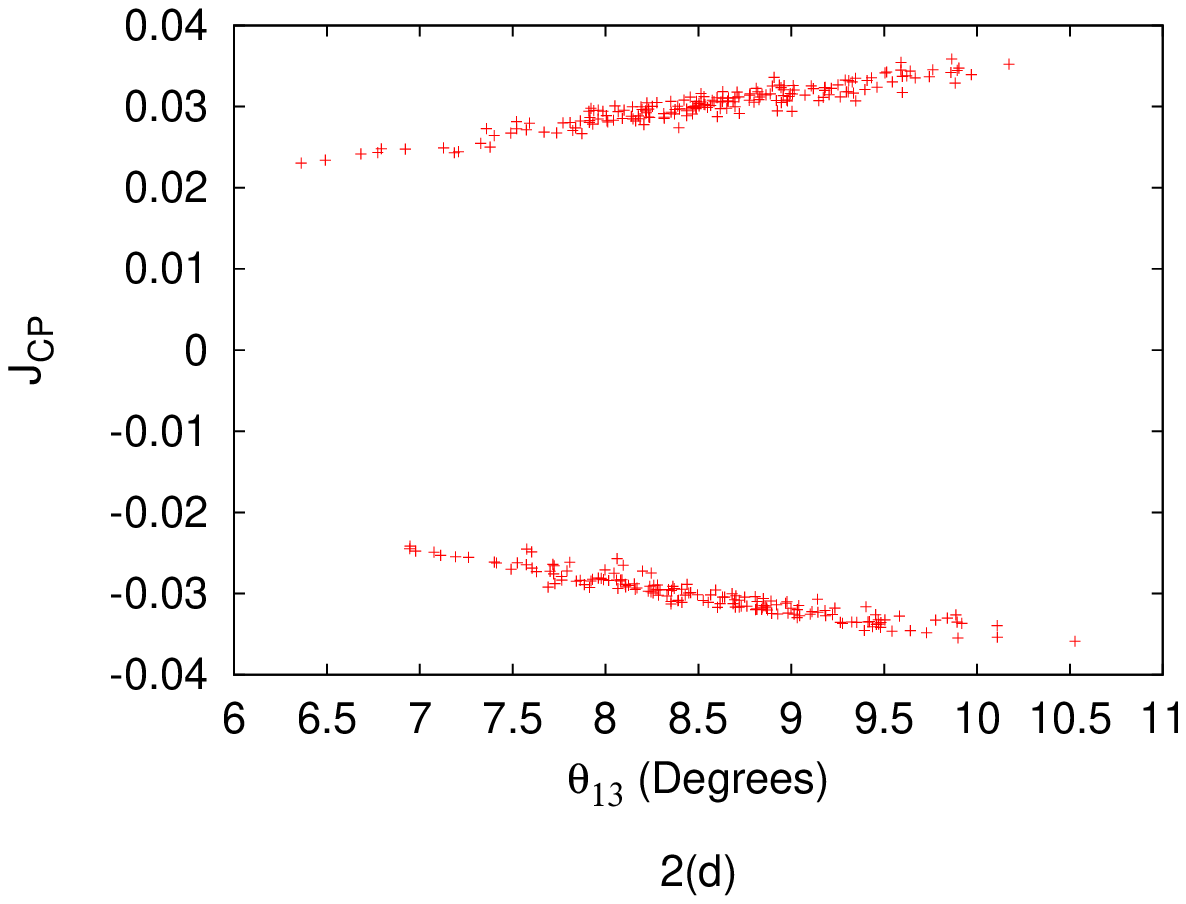, width=7.0cm, height=7.0cm}}

\caption{The correlation plots of $\kappa$, $\theta_{13}$ and $J_{CP}$ for inverted hierarchy (IH) of neutrino masses. The left (right) panel is without (with) Daya Bay result for $\theta_{13}$.}
\end{center}
\end{figure}

\begin{figure}
\begin{center}
{\epsfig{file=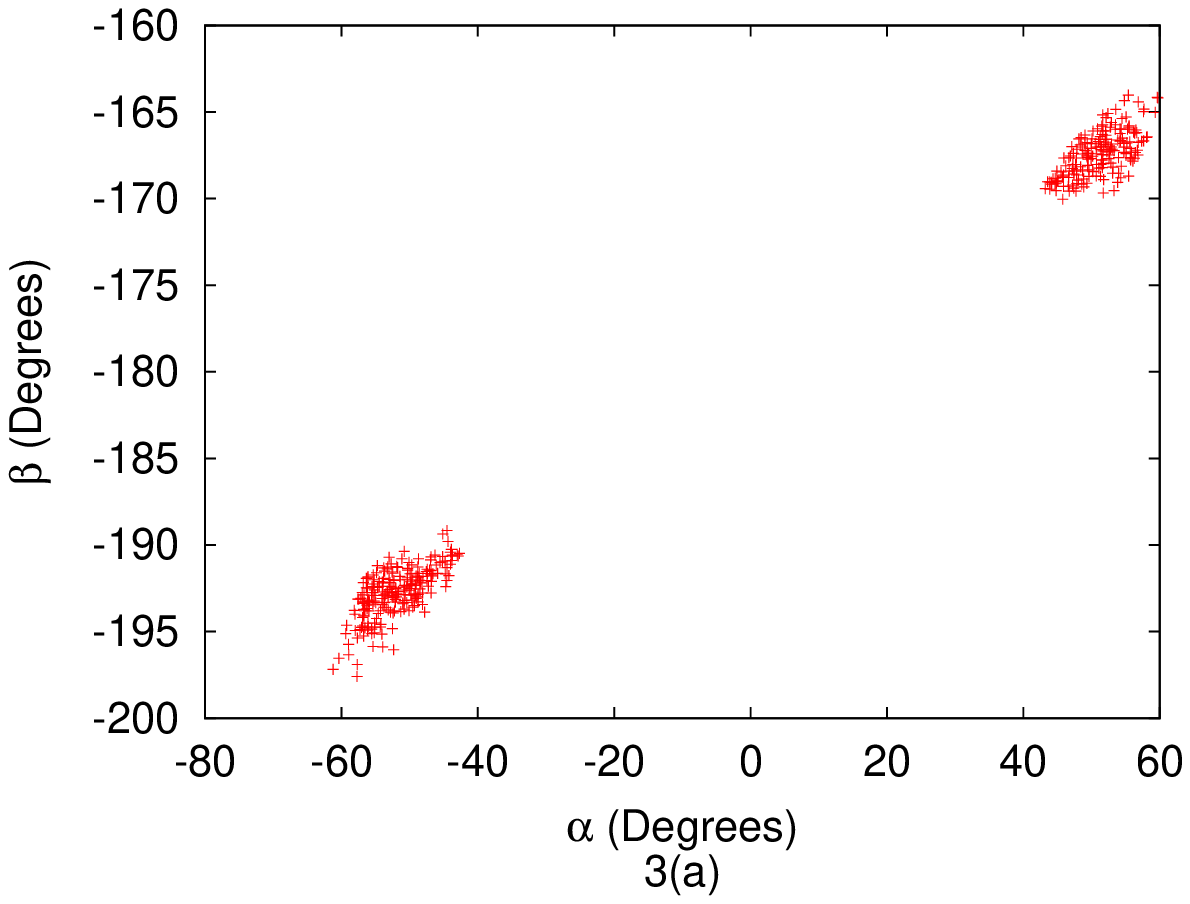, width=7.0cm, height=7.0cm} 
 \epsfig{file=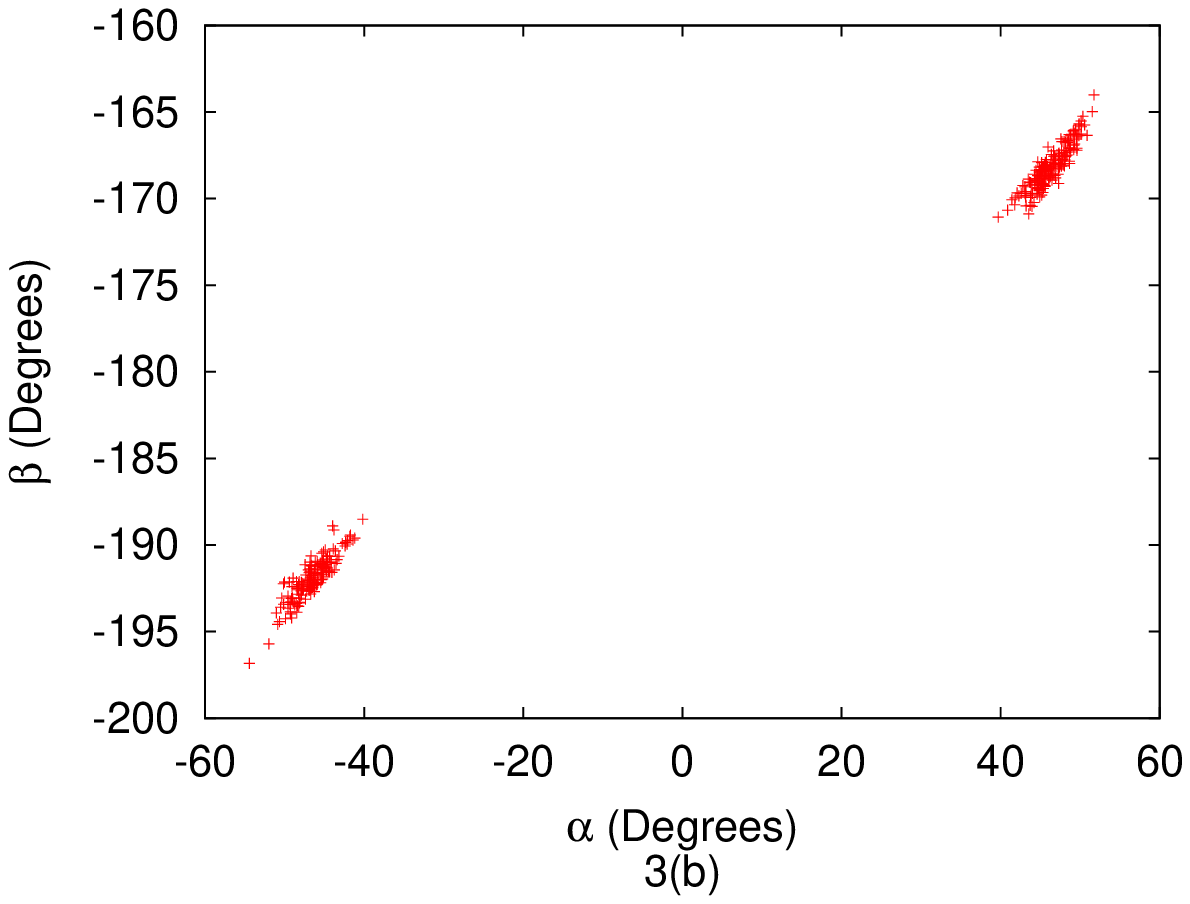, width=7.0cm, height=7.0cm}}
 {\epsfig{file=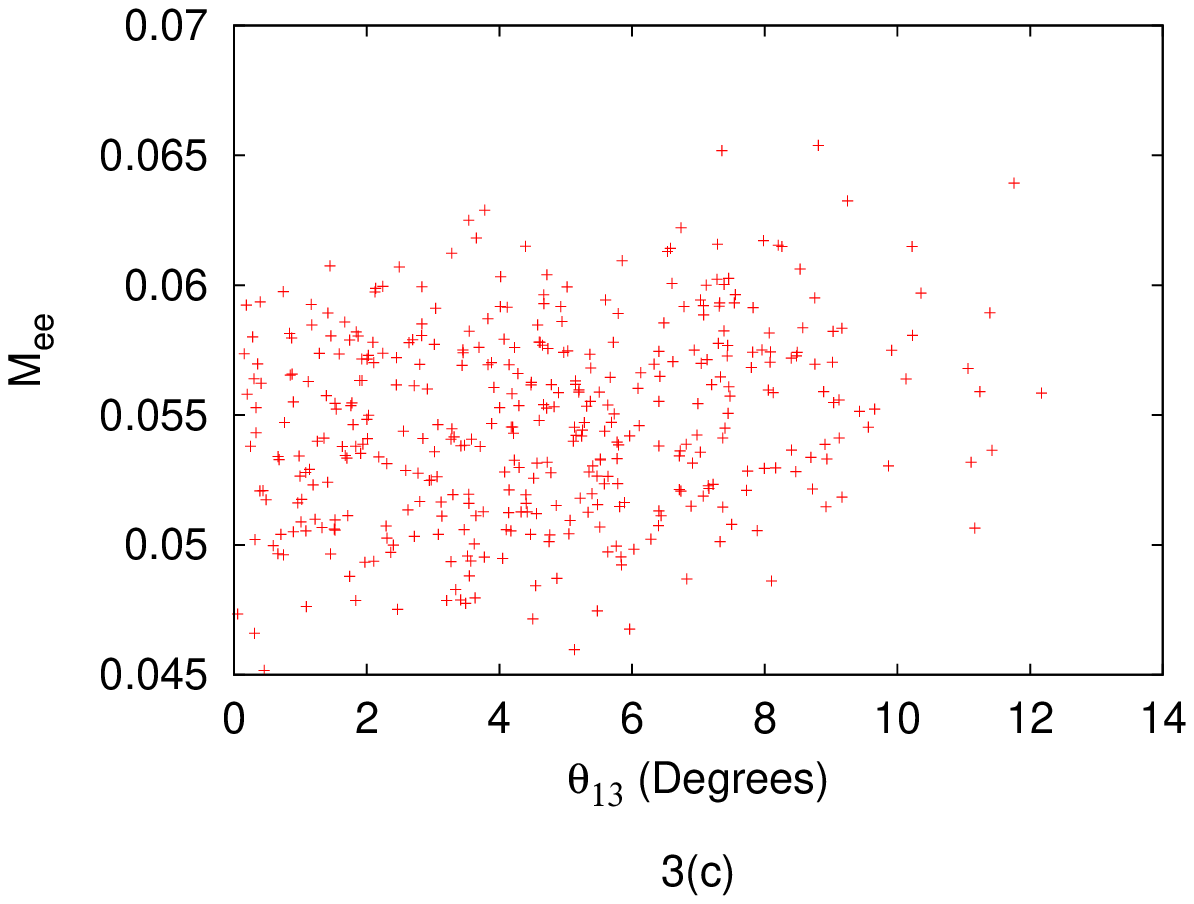, width=7.0cm, height=7.0cm} 
 \epsfig{file=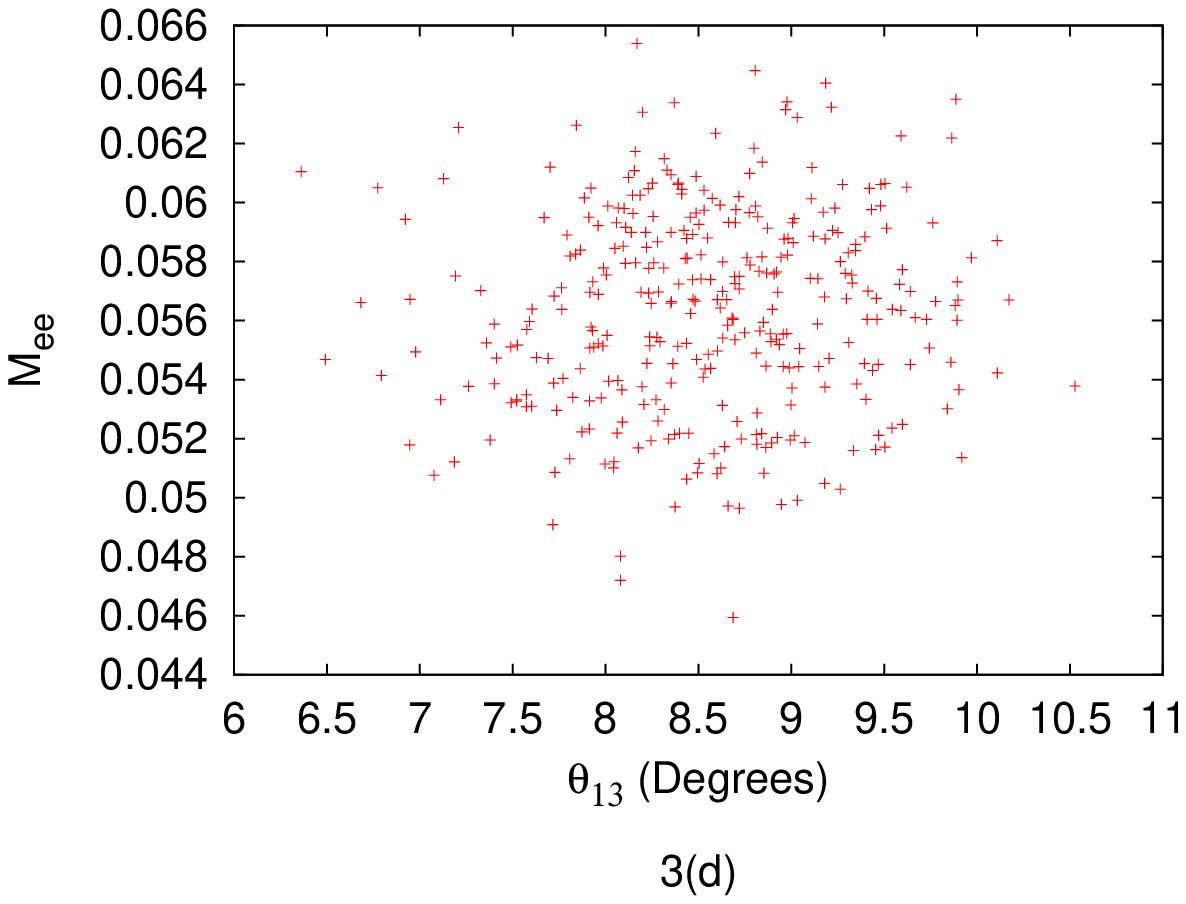, width=7.0cm, height=7.0cm}}
 
\caption{The correlation plots of Majorana phses $\alpha$, $\beta$, $\theta_{13}$ and effective Majorana neutrino mass $M_{ee}$(eV) for inverted hierarchy (IH) of neutrino masses.  The left (right) panel is without (with) Daya Bay result for $\theta_{13}$.}
\end{center}
\end{figure}


\begin{thebibliography}{99}
\bibitem{1} F. P. An \textit{et al.} [Daya Bay Collaboration], ``Observation of electron-antineutrino disappearance at Daya Bay" [arXiv:1203.1669[hep-ex]].
\bibitem{2} K. Abe \textit{et al.} [T2K Collaboration], \textit{Phys. Rev. Lett.} \textbf{107}, 041801 (2011) [arXiv:1106.2822[hep-ex]].
\bibitem{3} P. Adamson \textit{et al.} [MINOS Collaboration], \textit{Phys. Rev. Lett.} \textbf{107}, 181802 (2011) [arXiv:1108.0015[hep-ph]].
\bibitem{4} H. De Kerret \textit{et al.} [DOUBLE CHOOZ Collaboration], talk presented at the Sixth International Workshop on Low Energy Neutrino Physics, November 9-11, 2011 (Seoul, Korea).
\bibitem{5} T. Schwetz, M. Tortola and J. W. F. Valle, \textit{New J. Phys.} \textbf{13}, 109401 (2011) [arXiv:1108.1376[hep-ph]]; M. C. Gonjalez-Garcia, M. Maltoni and J. Salvado, \textit{JHEP} \textbf{1004}, 056 (2010) [arXiv:1001.4524v3[hep-ph]]; G. L. Fogli, E. Lisi, A. Marrone, A. Palazzo and A. M. Rotunno, [arXiv:1106.6028[hep-ph]].
\bibitem{6} Surender Verma, \textit{Nucl. Phys.} \textbf{B 854}, (2012) 340-349 [arXiv:1109.4228v1[hep-ph]]; Y. H. Ahn and Sin Kyu Kang, [arXiv:1203.4185v1[hep-ph]]; D. Meloni, S. Morisi and E. Peinado, [arXiv:1203.2535[hep-ph]]; Xinyi Zhang and Bo-Qiang Ma, [arXiv:1203.2906v1[hep-ph]]; H. Fritzsch, [arXiv:1203.4460v1[hep-ph]]; Shu Luo, Zhi-zhong Xing, [arXiv:1203.3118v1[hep-ph]]; G. C. Branco, R. Gonzalez Felipe, F. R. Joaquim, H. Serodio, [arXiv:1203.2646v1[hep-ph]]; Hong-Jian He and Xun-Jie Xu, [arXiv:1203.2908v1[hep-ph]]; Zhi-zhong Xing, [arXiv:1203.1672v1[hep-ph]].
\bibitem{7} R. N. Mohapatra and W. Rodejohann, \textit{Phys. Lett.} \textbf{B 644}, (2007) 59-66; W. Grimus and L. Lavoura, \textit{J. Phys.} \textbf{G 31}, 683 (2005) [arXiv:hep-ph/0410279]; W. Grimus and L. Lavoura, \textit{Phys. Rev.} \textbf{D 62}, 093012 (2000) [arXiv:hep-ph/0007011]; L. Lavoura, \textit{Phys. Rev.} \textbf{D 62}, 093011 (2000) [arXiv:hep-ph/0005321]; A. S. Joshipura and W. Rodejohann [arXiv:0905.2126[hep-ph]].
\bibitem{8} P. Minkowski, \textit{Phys. Lett.} \textbf{B 67}, 421 (1977); T. Yanagida, in \textit{Proceedings of the Workshop on the Unified Theory and the Baryon Number in the Universe} (O. Sawada and A. Sugamoto, eds.), KEK, Tsukuba, Japan, 1979, p. 95; M. Gell-Mann, P. Ramond, and R. Slansky, \textit{Complex spinors and unified
theories}, in \textit{Supergravity} (P. van Nieuwenhuizen and D. Z. Freedman, eds.), North Holland, Amsterdam, 1979, p. 315; S. L. Glashow, \textit{The future of elementary particle physics}, in \textit{Proceedings of the 1979 Cargese Summer Institute on Quarks and Leptons} (M. Levy, J. L. Basdevant, D. Speiser, J. Weyers, R. Gastmans and M. Jacob, eds.), Plenum Press, Newyork, 1980, pp. 687-713; R. N. Mohapatra and G. Senjanovic, \textit{Phys. Rev. Lett.}  \textbf{44}, 14 (1980).
\bibitem{9} Midori Obara, [arXiv:0712.2628v2[hep-ph]].
\bibitem{10} A. Blum, R. N. Mohapatra and W. Rodejohann, \textit{Phys. Rev.} \textbf{D 76}, 053003 (2007) [arXiv:0706.3801[hep-ph]].
\bibitem{11} K. Nakamura \textit{et al.} [Particle Data Group], \textit{J. Phys.} \textbf{G 37}, 075021 (2010) and 2011 partial update for the 2012 edition.
\bibitem{12} C. Jarlskog, \textit{Phys. Rev. Lett.} \textbf{55}, 1039 (1985).
\bibitem{13} H. V. Klapdor-Kleingrothaus, A. Dietz, H. L. Harney and I. V. Krivosheina, \textit{Mod. Phys. Lett.} \textbf{A 16}, (2001) 2409; H. V. Klapdor-Kleingrothaus, I. V. Krivosheina, A. Dietz and O. Chkvorets, \textit{Phys. Lett.} \textbf{B 586}, (2004) 198; S. M. Bilenky, Amand Faessler, W. Potzel and F. Simkovic, [arXiv:1104.1952[hep-ph]].  
\bibitem{14} C. Arnaboldi \textit{et al.}, \textit{Nucl. Instr. Meth.} \textbf{A 518}, 775 (2004); I.C. Bandac \textit{et al.}, \textit{J. Phys. Conf. Ser.} \textbf{110}, 082001 (2008).
\bibitem{15} J. Janicsko-Csathy, \textit{Nucl. Phys.} \textbf{B 188}, 68 (2009) (Proc. Suppl.); S.R. Elliott \textit{et al.}, \textit{J. Phys. Conf. Ser.} \textbf{173}, 012007 (2009).
\bibitem{16} A.S. Barabash, \textit{Czech. J. Phys.} \textbf{52}, 575 (2002); A.S. Barabash, \textit{Phys. At. Nucl.} \textbf{67}, 1984 (2004).
\bibitem{17} M. Danilov \textit{et al.}, \textit{Phys. Lett.} \textbf{B 480}, 12 (2000); R. Gornea, \textit{J. Phys. Conf. Ser.} \textbf{179}, 022004 (2009).


\end{thebibliography}
\end{document}